\documentclass[aps,prb,twocolumn,superscriptaddress,floatfix]{revtex4}
\usepackage{physics}
\usepackage{float}
\usepackage{graphicx}
\usepackage{color} 

\begin{document}
\title{Influence of State Reopening Policies in COVID-19 Mortality}
 
\author{Ka-Ming Tam}
\affiliation{Department of Physics \& Astronomy, Louisiana State University, Baton Rouge, Louisiana 70803, USA}
\affiliation{Center for Computation \& Technology, Louisiana State University, Baton Rouge, Louisiana 70803, USA}

\author{Nicholas Walker}
\affiliation{Department of Physics \& Astronomy, Louisiana State University, Baton Rouge, Louisiana 70803, USA}

\author{Juana Moreno}
\affiliation{Department of Physics \& Astronomy, Louisiana State University, Baton Rouge, Louisiana 70803, USA}
\affiliation{Center for Computation \& Technology, Louisiana State University, Baton Rouge, Louisiana 70803, USA}

\date{\today}

\begin{abstract}

By the end of May 2020, all states in the US have eased their COVID-19 mitigation measures. 
Different states adopted markedly different policies and timing for reopening. 
An important question remains in how the relaxation of mitigation measures is 
related to the number of casualties. To address this question, we compare the actual data to a hypothetical 
case in which the mitigation measures are left intact using a projection of the data from before mitigation measures were eased. We find that different states have shown significant differences in the number of deaths, possibly due to their different policies and reopening schedules. Our study provides a gauge for the effectiveness of the approaches by different state governments 
and can serve as a guide for implementing best policies in the future. It also indicates 
that the face mask mandate seems to correlate with the change in the death count more dramatically than other measures.

\end{abstract}

\maketitle 
\section{Introduction}


At of the beginning of November,  
there are close to 10 million confirmed cases and more than 237,000 casualties attributed to COVID-19 in the United States.  Although the first case was confirmed
on January 20, the number of reported cases was rather low until early March. 
The exceedingly slow spreading in these early months may have partially been due to the lack of adequate testing, which has remained a major issue for a long time.
The number of reported cases then dramatically increased in early March.

Since the growth rate of infections was alarmingly high in most areas by mid-March, the majority of states implemented mitigation efforts by the end of March. These measures included closing restaurants 
and bars, shutting down schools, and implementing stay-at-home orders. By early April, nearly all states were 
under some form of lock-down. 
The infections tapered down appreciably due to these mitigation efforts. The daily number of new cases seemed to peak in early April and both the number of new cases and deaths steadily decreased from April to late May. 

While the total number of infections never dropped below 16,000 per day, 
many states decided to reduce their mitigation efforts after this first wave.
By the end of May, all states had 
rolled back their restrictions in some capacity. 
Certain states were more aggressive in reopening than others. While the number of infections was not particularly large at the beginning of reopening since the observed effect of the spread is delayed due to the incubation time and waiting time to receive testing results, many states saw a sharp  
increase in the number of cases between late May and early June. This concerning 
situation forced several states to scale back plans for reopening in order to stabilize the exponentially increasing number of cases. 

There are several factors which are expected to reduce COVID-19 mortality rate. 
Adequate testing was rather limited before May. Testing has since expanded appreciably and stabilized at about 800,000 per day. One would expect an increase in testing rate to reduce the number of fatalities, as people who 
tested positive as well as their close contacts should be quarantined. 
Moreover, the death rate was particularly high at the beginning of the pandemic 
due to the large number of
infections in high risk groups, such as residents of assisted living facilities.  
The fact that the average age of infected persons has been decreasing since March should 
correspond to a reduction in the death rate. 
Finally, improvement in treatments should also help to reduce the mortality rate. Together with the awareness of preventative measures, such as wearing masks in public areas, all of these factors should help to reduce the mortality rate and the number of casualties per day providing that 
the mitigation efforts remain intact. 
 
Since all of these factors point to a decrease in the number of fatalities, 
the marked increase in the number of deaths since late May clearly indicates that relaxation of the mitigation efforts is the prominent reason for the increase. This is
also consistent with the fact that states which were more aggressive in reopening, in terms of both the timing and the restrictions lifted, seem 
to have experienced a higher percentage increase in death counts. Therefore we believe understanding the relation between reopening policies and  death count is a timely topic. 

This paper provides an estimate of the change 
in the number of fatalities with respect to easing mitigation efforts. We
not only quantify the additional number of deaths, but more importantly, by comparing between different states, this study provides clues to the reopening strategy which could minimize loss of life.
We find significant differences between different states, as while most states display an increase in the number
of fatalities, a handful of states actually show a decrease. This should be important and timely information as another infection wave may occur in the winter season. 

This paper is organized as follows. In Section II, we describe the model and methods. The results are presented Section III. In section IV, we discuss our
results in the context of the different state reopening policies. We conclude our discussion in Section V.


\section{Model} \label{Sec:Model}

We employ the Susceptible-Infected-Recovered (SIR) model\cite{Huppert,Kermack_McKendrick} modified to consider the number of quarantined people. Similar modifications on the SIR model have been considered elsewhere to model the spread of COVID-19.\cite{Crokidakis,Bin,Pedersen,Calafiore,Bastos,Gaeta1,Gaeta2,Vrugt,Schulz,Zhang,Amaro,DellAnna,Sonnino,Notari,Simha,Acioli,Zullo,Sameni,Radulescu,Roques,Teles,Piccolomini,Brugnano,Giordano,Zlatic,Baker,Biswas,Zhang_Wang_Wang,Chen,Lloyd,Vadyala,Fokas1,Fokas2,Cooper,Bertozzi,Prem,He,Mwalili,Dehning,Atkeson,Wang} 
This model is essentially a coarse grained or mean field approach that assumes the population is homogeneous and well mixed and that the dynamics do not exhibit an explicit time delay. The advantage of this model is in its simplicity, as it tries to describe the development of an epidemic with a minimal number of variables and parameters. It allows for an intuitive understanding of the spreading of the virus, which is not often the case for a generic model of time series with a lot of tunable parameters. The limitations of the model are thoroughly discussed in a recent publication. \cite{Tam_Walker_Moreno_2020b} The equations defining the dynamics of the model are as follows:

\begin{align}
\dv{S(t)}{t} &= -\beta \frac{S(t)I(t)}{N},\label{Eq:Seq}\\
\dv{I(t)}{t} &= \beta \frac{S(t)I(t)}{N} -\qty(\alpha +\eta)I(t),\label{Eq:Ieq}\\
\dv{Q(t)}{t} &= \eta I(t)-\delta(t) Q(t) - \xi(t) Q(t),\label{Eq:Qeq}\\
\dv{R(t)}{t} &= \xi(t) Q(t) +\alpha I(t),\label{Eq:Req}\\
\dv{C(t)}{t} &= \delta(t) Q(t), \label{Eq:Ceq}
\end{align}
where $N$ is the total population size, $S$ is the susceptible population count, $I$ is the unidentified while infectious population count, $Q$ is the number of identified positive cases which are quarantined,  $R$ is the number of recovered patients, and $C$ is the number of deaths.
The model is characterized by the following parameters: $\beta$ is the infection rate, $\eta$ is the detection rate, $\alpha$ is the recovery rate of asymptomatic people, $\xi$ is the recovery rate of the quarantined patients, and $\delta$ is the casualty rate of the quarantined. 
All of the parameters are in units of (1/day). The quarantined population $Q$ is composed of the identified positive cases regardless of whether they are hospitalized or at home.
We further assume that all casualties had been in quarantine prior to death and we consider that only $\xi$ and $\delta$ are time dependent out of the coefficients. All of these assumptions are approximations made to allow for inference of the model parameters from the currently available data.




The total death count at time $t$,  $D(t)$,  can be estimated as:

\begin{equation}
    D(t) = \int_{0}^{t} \dv{C(\tau)}{\tau} \dd{\tau}.
\end{equation}

The confirmed positive count is $P(t)= Q(t)+R_Q(t)+C(t)$, where $R_Q(t)$ are the recovered patients previously in quarantine. $P(t)$ can be estimated as:

\begin{eqnarray}
    P(t)&=& \int_{0}^{t} \dv{P(\tau)}{\tau} \dd{\tau} \\&=& \int_{0}^{t} 
    \left( \dv{Q(\tau)}{\tau} +\dv{R_Q(\tau)}{\tau}+\dv{C(\tau)}{\tau}\right) \dd{\tau}, \nonumber \\
    &=& \int_{0}^{t} \eta I(\tau) \dd{\tau}.
    \label{Eq:P(t)}
\end{eqnarray}

\section*{Method} \label{Sec:Method}

We determine two sets of parameters, one before the stay-at-home order and the other after the social distancing measures are in place, as 
discussed in our previous work.~\cite{Tam_Walker_Moreno_2020a,Tam_Walker_Moreno_2020b} 
Instead of extrapolating the data from other areas or countries, we choose to determine it from the casualty and confirmed case counts in each state separately.
We estimate the first set of parameters by fitting the data at the beginning of the epidemic to an exponential growth. We consider the effects of social distancing measures to be reflected in two parameters, the reduction of the infection rate, 
and the first day when the measurements are effective, 
since there is a time delay in the influence the stay-at-home orders have on the number of cases and deaths.
We determine both parameters by minimizing the $\chi^2$ of the values and daily changes of the death and the confirmed infected count based on the data before April 30th.~\cite{Tam_Walker_Moreno_2020b}  
The data for the daily casualties and new cases are obtained from the database of the New York Times. \cite{NYT_data}

\section{Results}

\begin{figure*}
    \begin{minipage}{.45\textwidth}
        \centering
        \includegraphics[width=\textwidth]{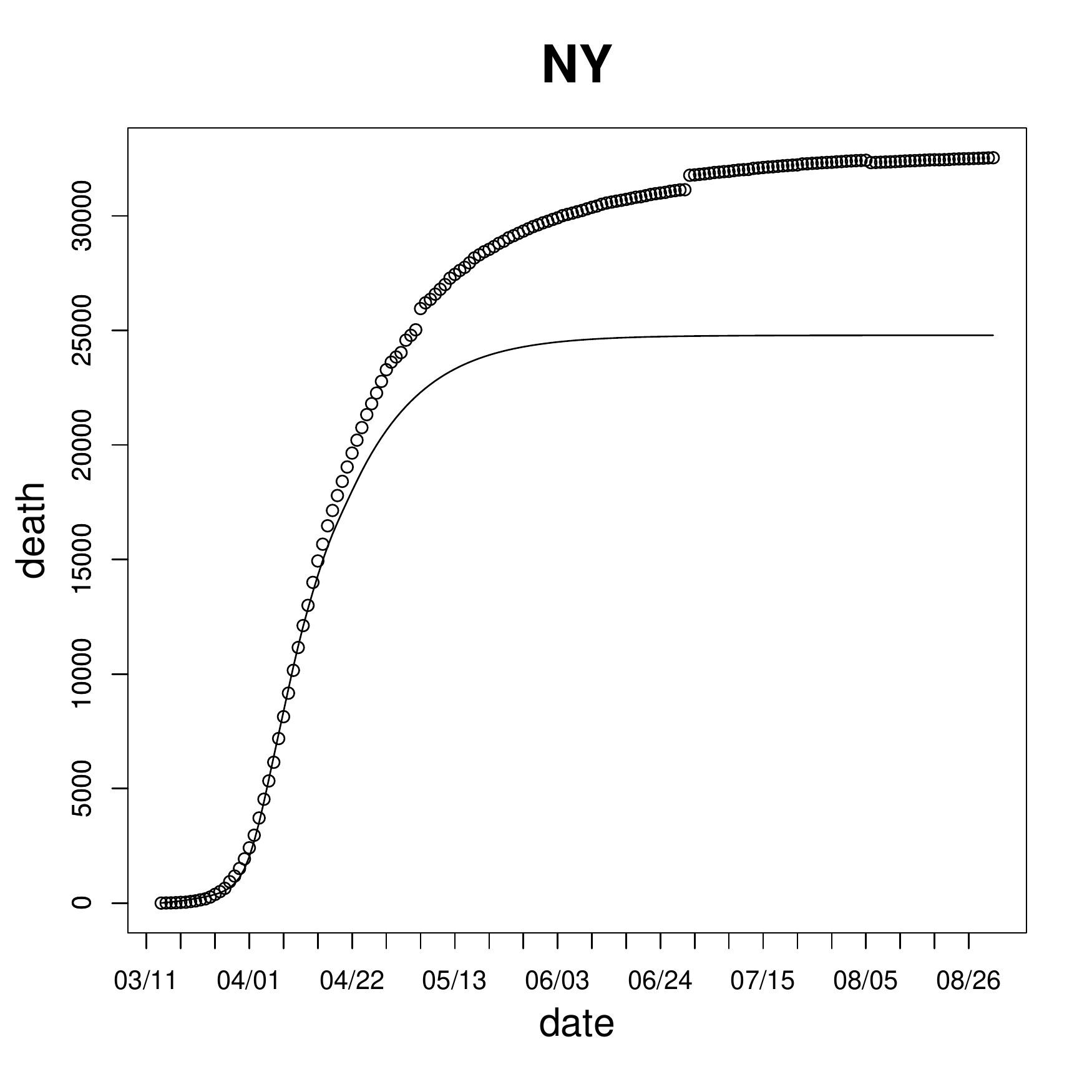}
    \end{minipage}
   \hfill
    \begin{minipage}{.45\textwidth}
        \centering
        \includegraphics[width=\textwidth]{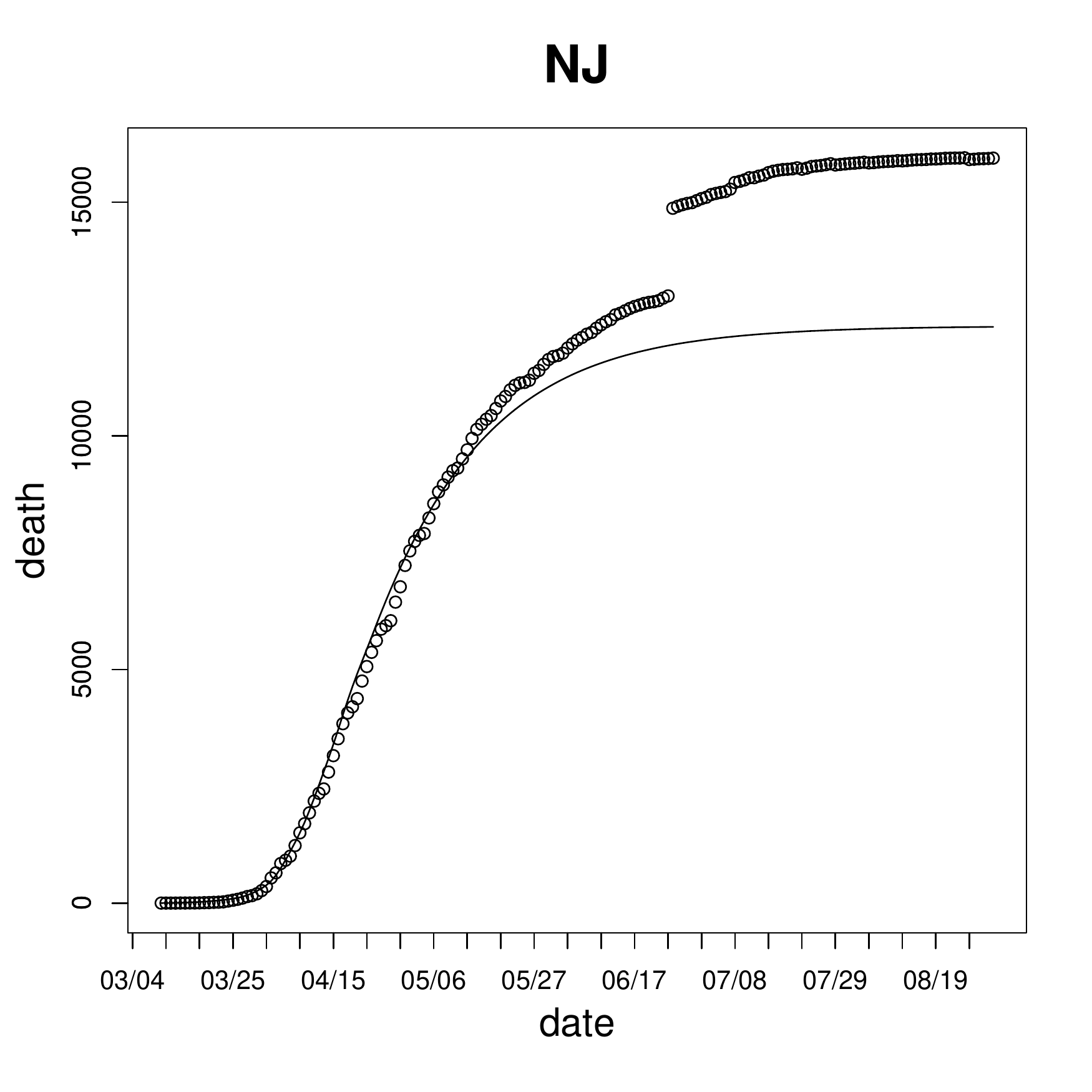}
    \end{minipage}
      \hfill
    \begin{minipage}{.45\textwidth}
        \centering
        \includegraphics[width=\textwidth]{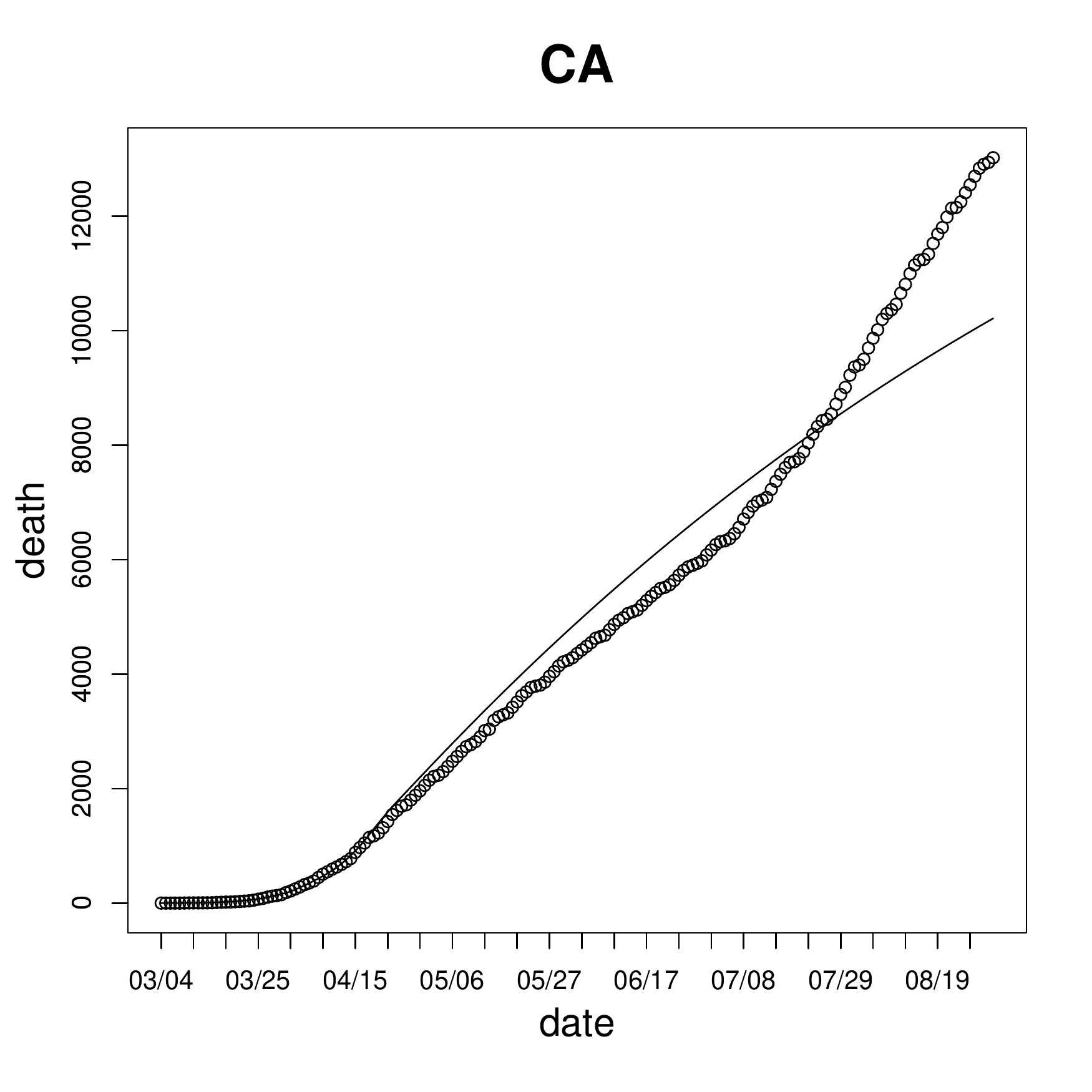}
    \end{minipage}  
       \hfill
    \begin{minipage}{.45 \textwidth}
        \centering
        \includegraphics[width=\textwidth]{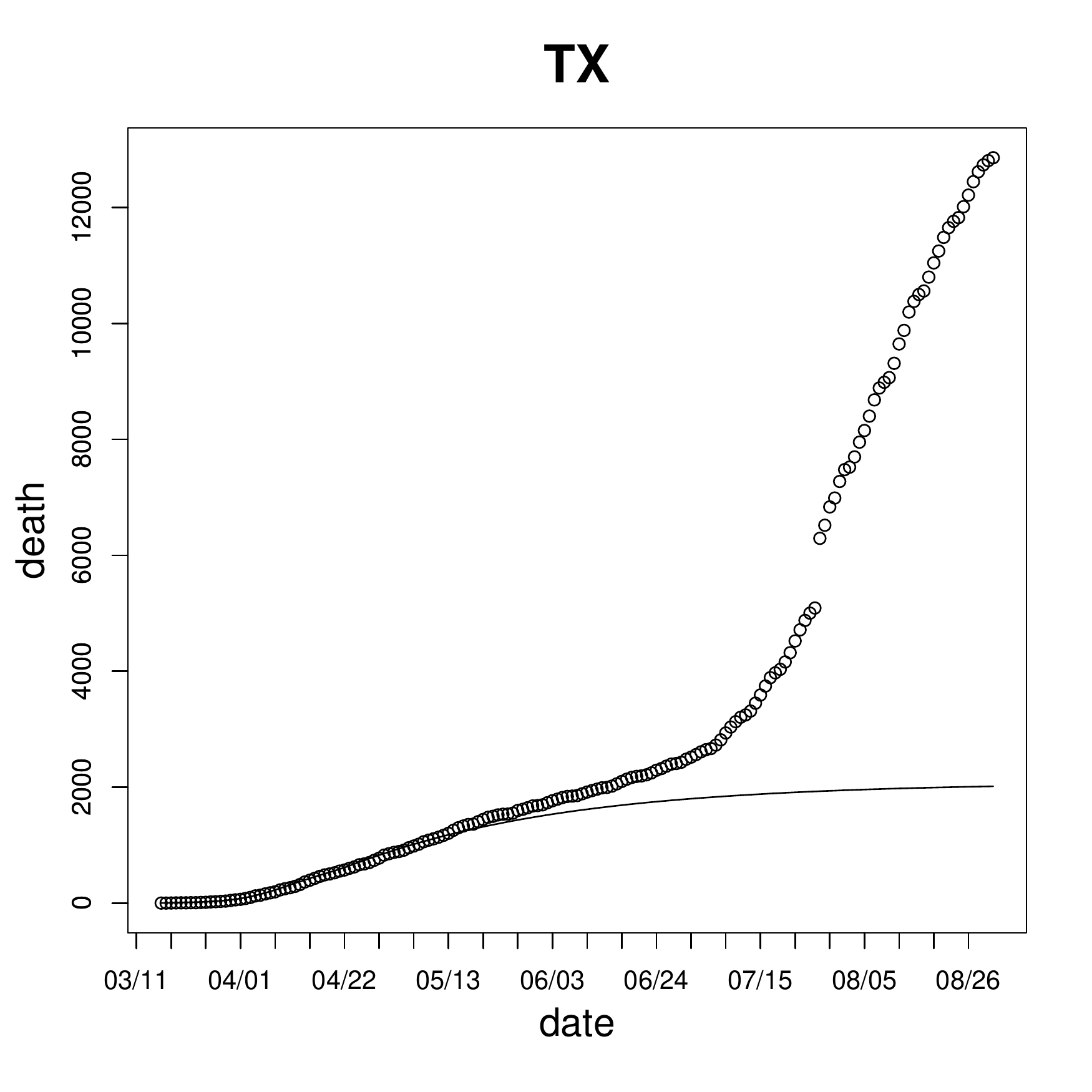}
    \end{minipage}  
    \begin{minipage}{.45\textwidth}
        \centering
        \includegraphics[width=\textwidth]{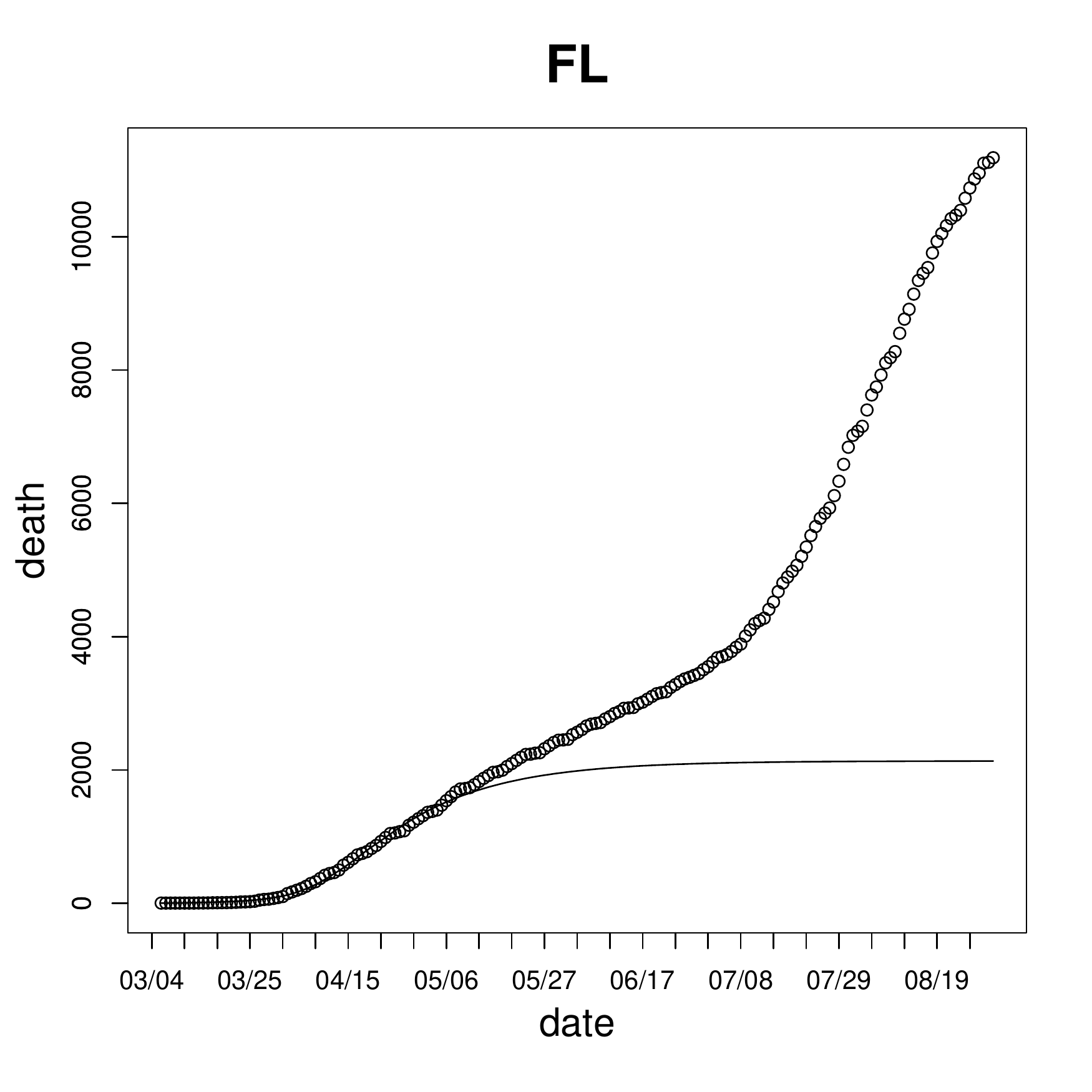}
    \end{minipage} 
         \hfill
    \begin{minipage}{.45\textwidth}
        \centering
        \includegraphics[width=\textwidth]{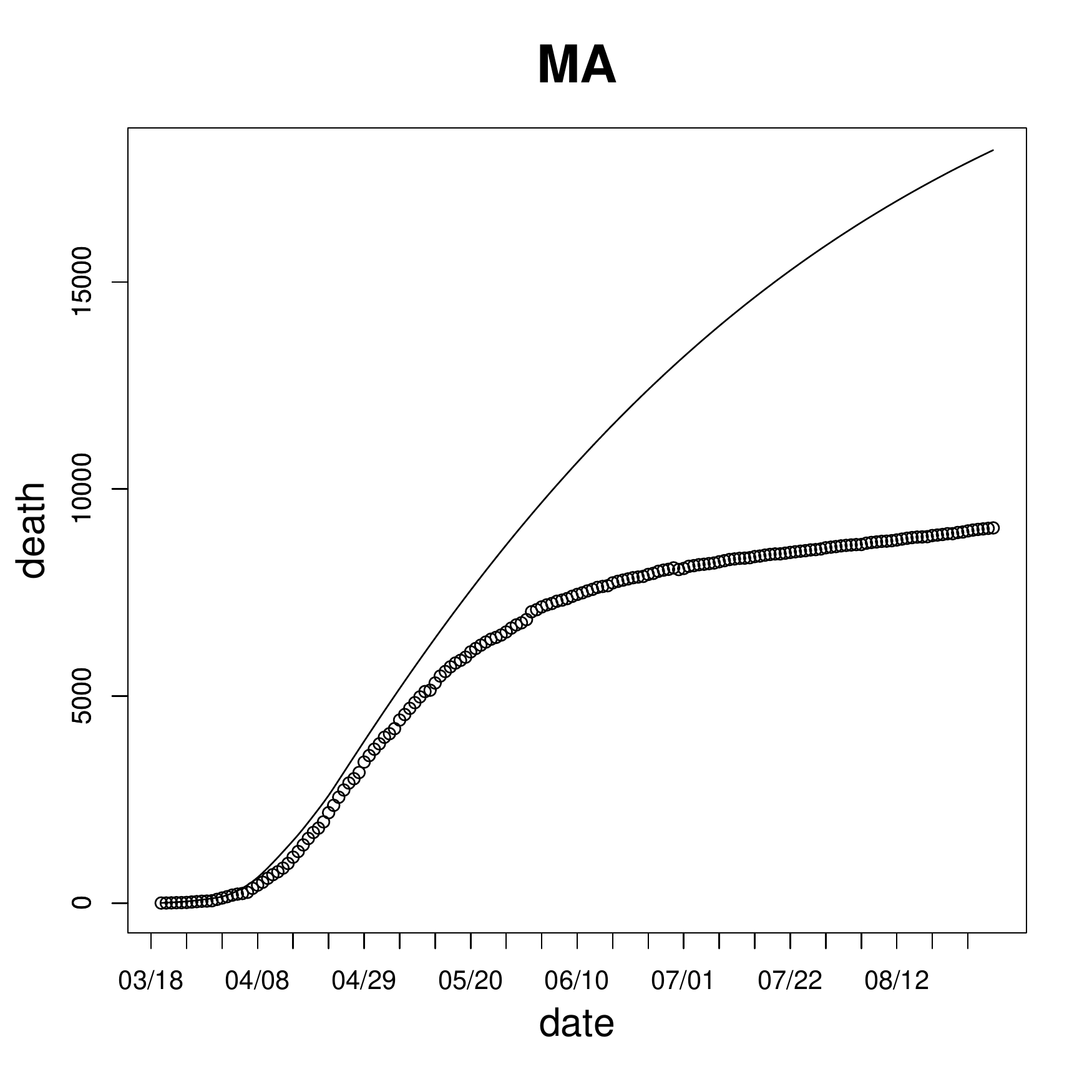}
    \end{minipage}
    \label{Fig:Comapre1}
   \caption{ Total death count (dots) \cite{NYT_data} and projected death count without easing the mitigation (solid line)
    as functions of time for the states of New York, New Jersey,  California, Texas, Florida, and Massachusetts using estimate by the data before the end of April. } 

\end{figure*}

\begin{figure*}
    \begin{minipage}{.45\textwidth}
        \centering
        \includegraphics[width=\textwidth]{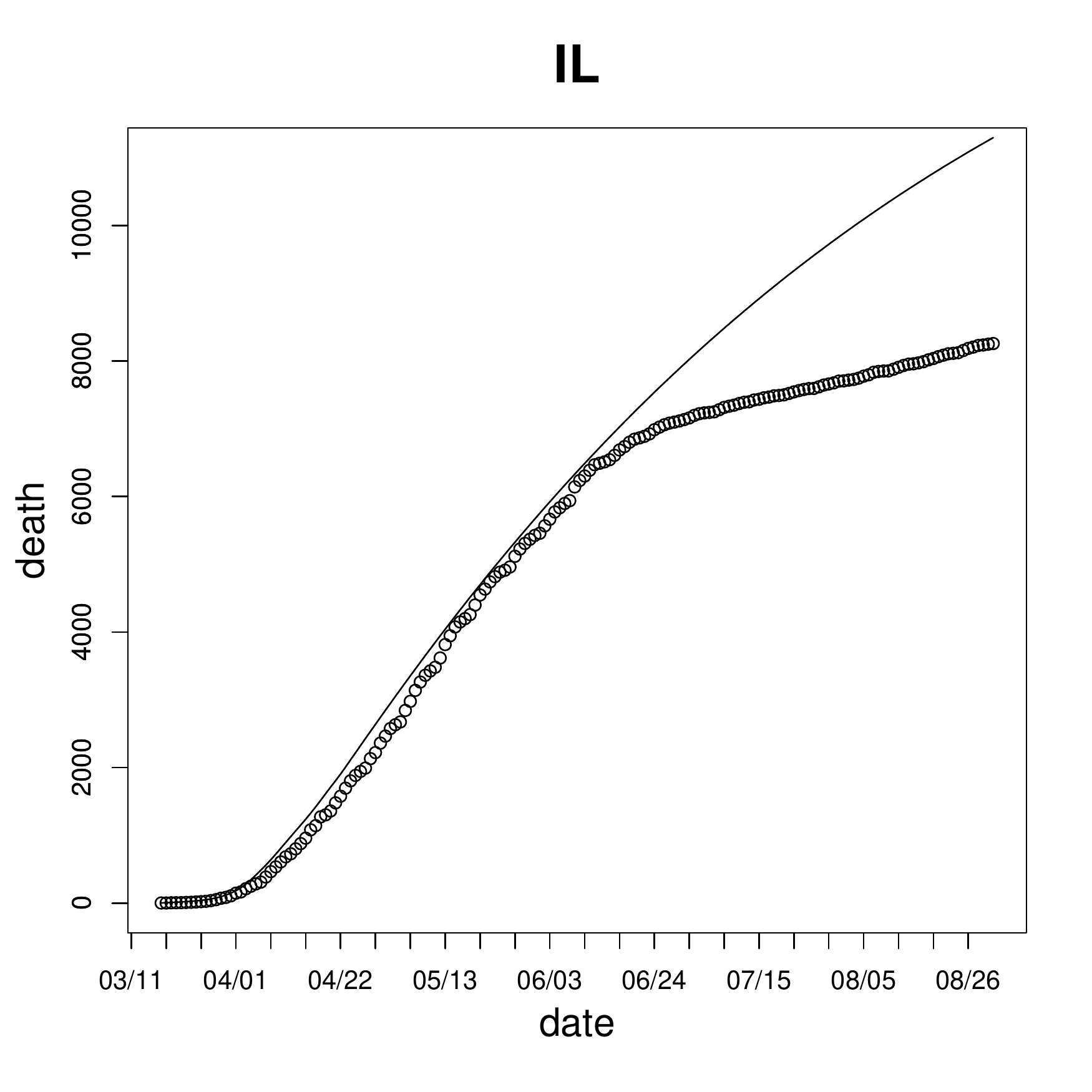}
    \end{minipage}
   \hfill
    \begin{minipage}{.45\textwidth}
        \centering
        \includegraphics[width=\textwidth]{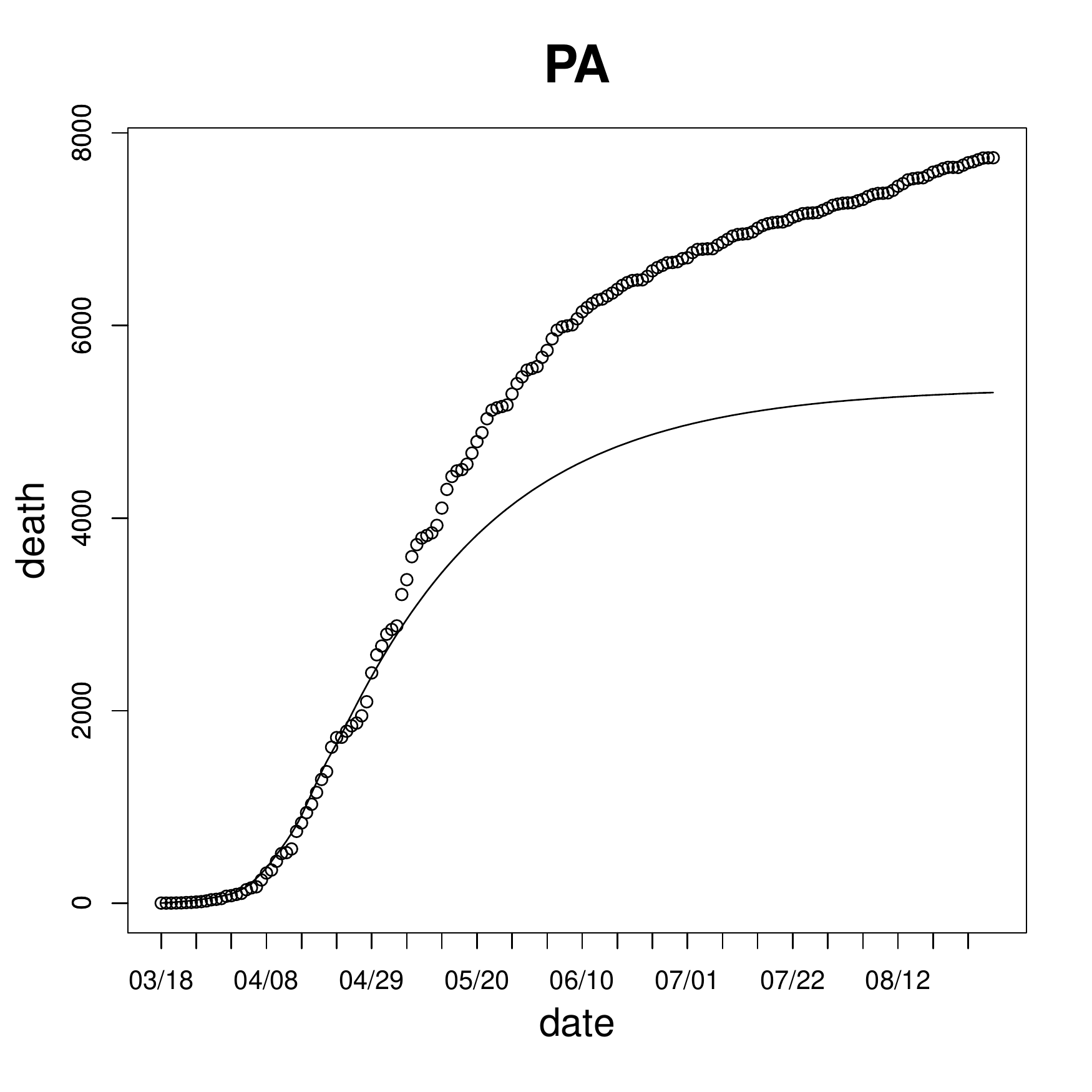}
    \end{minipage}
       \hfill
    \begin{minipage}{.45\textwidth}
        \centering
        \includegraphics[width=\textwidth]{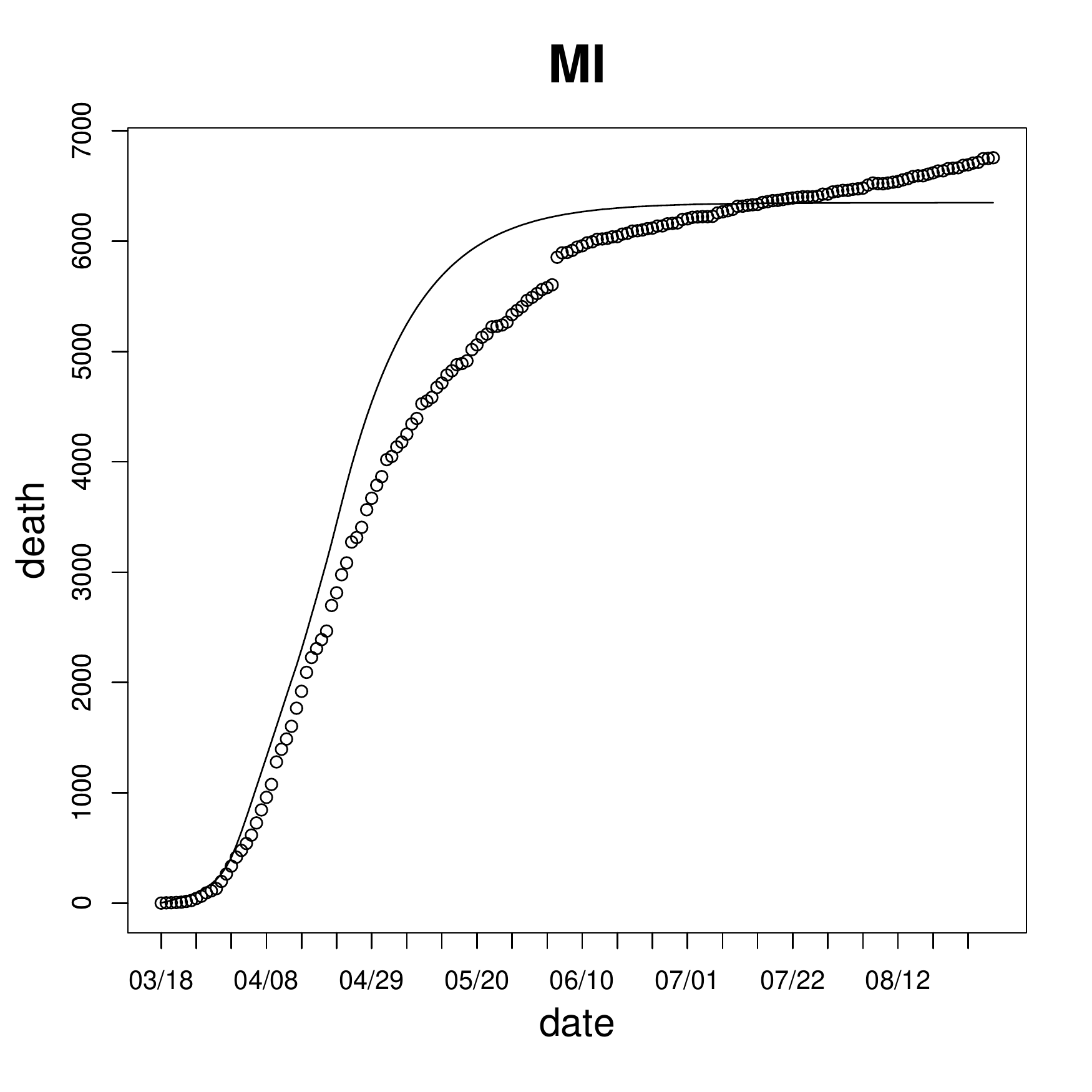}
    \end{minipage}  
       \hfill
    \begin{minipage}{.45 \textwidth}
        \centering
        \includegraphics[width=\textwidth]{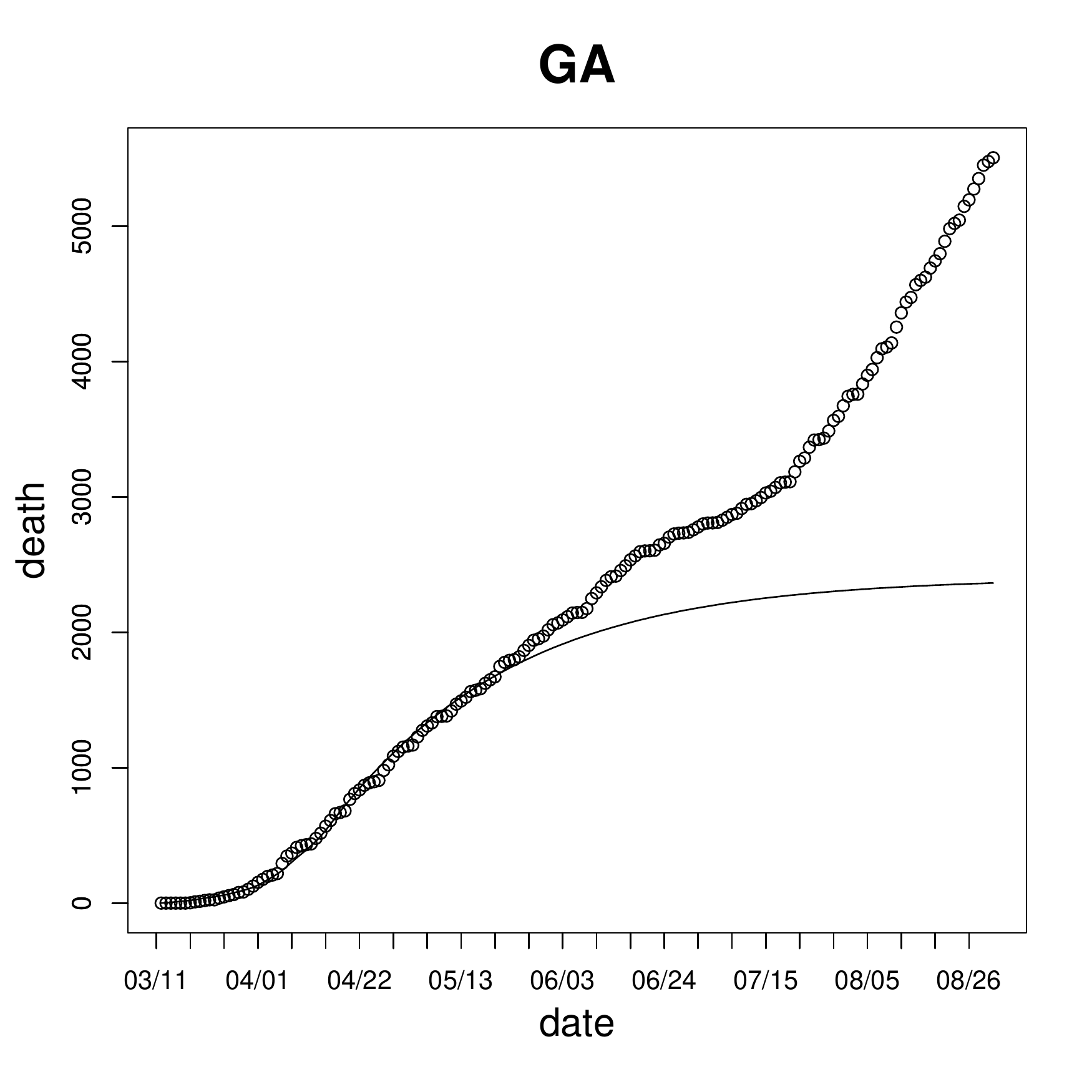}
    \end{minipage}  
    \begin{minipage}{.45\textwidth}
        \centering
        \includegraphics[width=\textwidth]{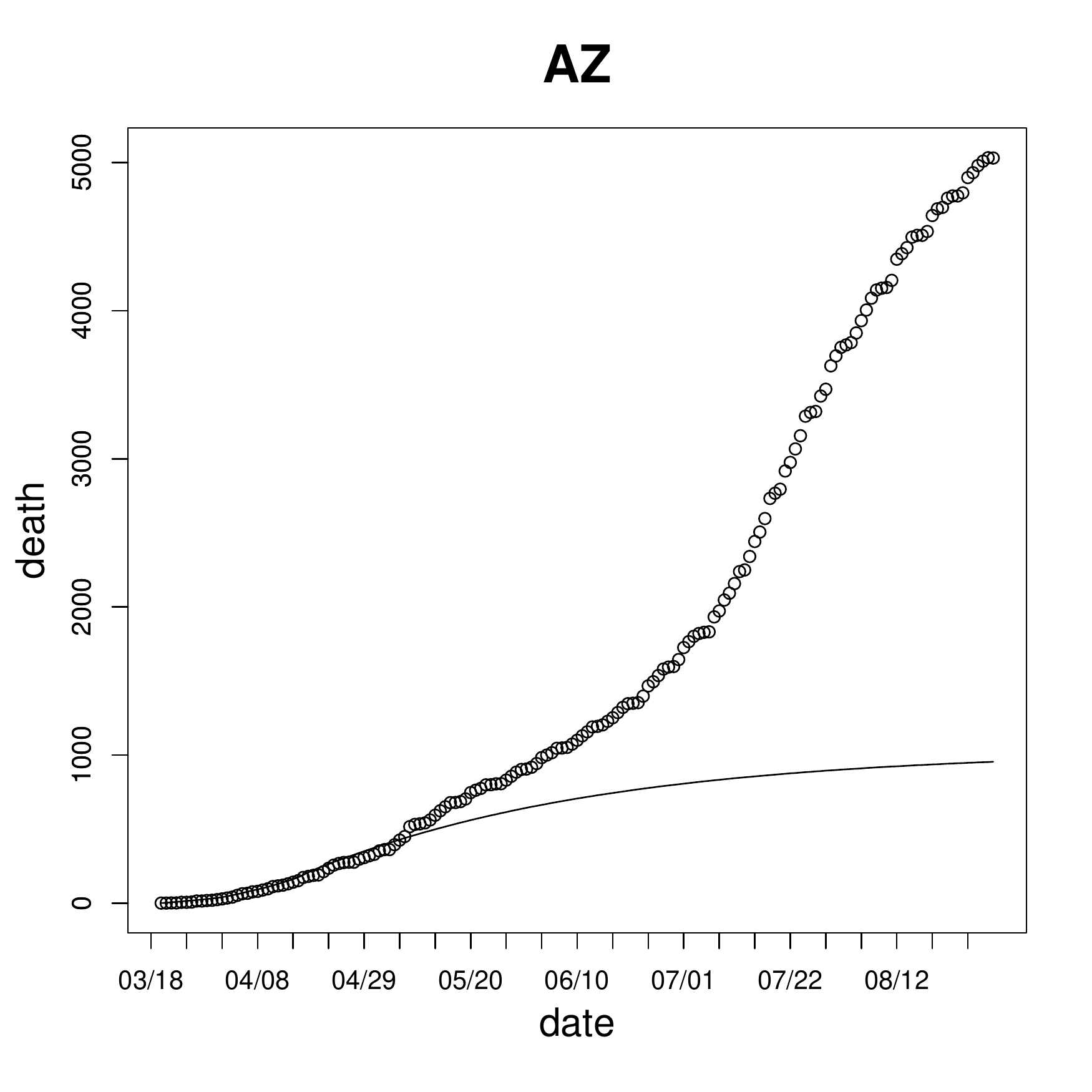}
    \end{minipage} 
         \hfill
    \begin{minipage}{.45\textwidth}
        \centering
        \includegraphics[width=\textwidth]{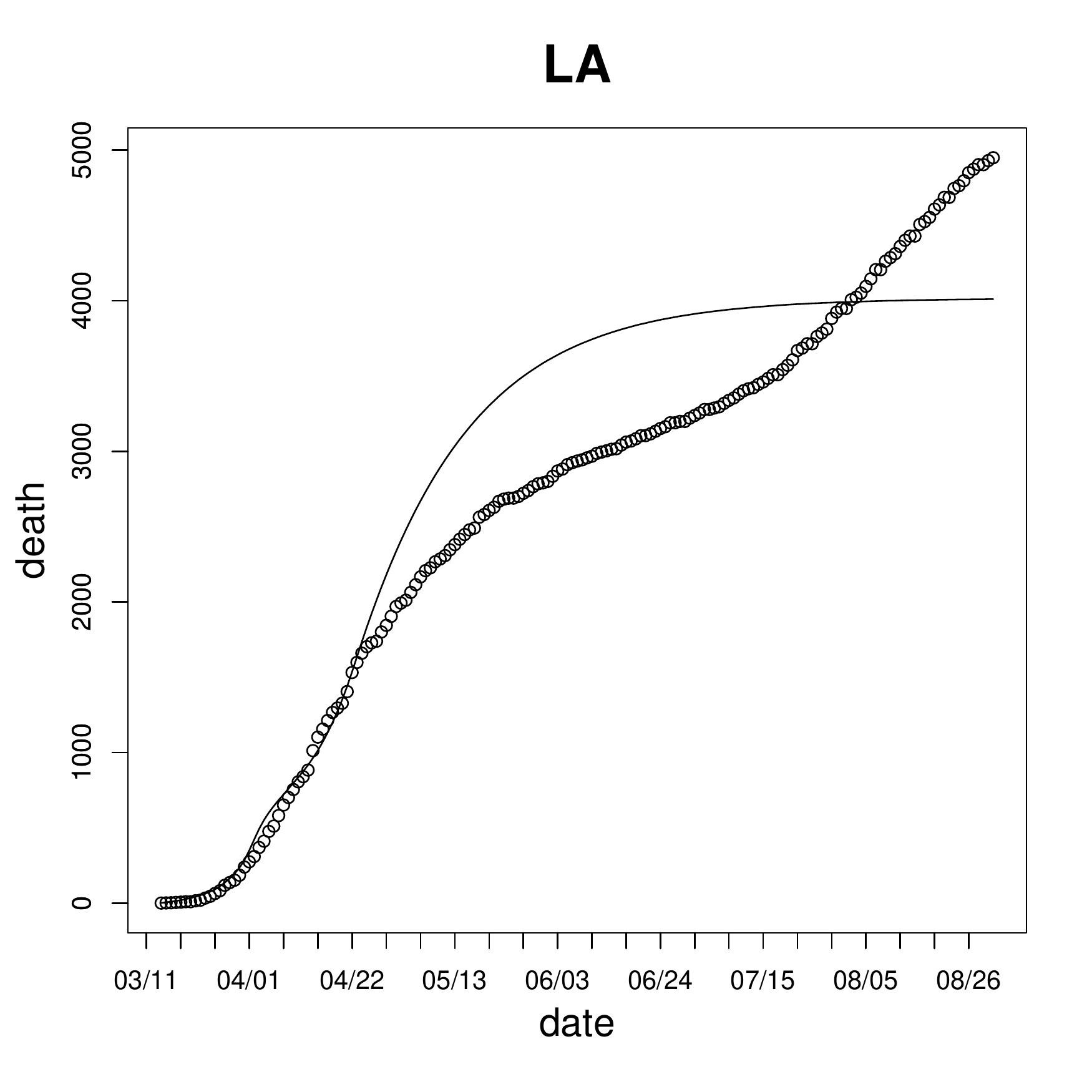}
    \end{minipage}  
    \label{Fig:compare2}
    \caption{ Total death count (dots) \cite{NYT_data} and projected fatalities count without easing the mitigation efforts (solid line)
    as functions of time for the states of
    Illinois, Pennsylvania, Michigan, Georgia, Arizona, and Louisiana using estimate by the data before the end of April.} 
    
\end{figure*}

Within the present model, there are two major routes for slowing the initial exponential growth of the epidemic, 
which is characterized  by the parameter $\beta -(\alpha + \eta)$. 
The first route is to decrease the infection rate $\beta$ and the second route is to increase the testing rate $\eta$. Mitigation efforts in the spring sharply reduced the infection rate.~\cite{Tam_Walker_Moreno_2020b} Testing 
rate also grew during the spring and summer. \cite{Atlantic_data}
Increasing the recovery rate of unidentified infected people, $\alpha$, can also reduce the spread, but this is more difficult to achieve. Overall, we can expect that the recovery rate should improve slightly over time. 
The average age of an infected person has trended towards decreasing since March. \cite{Atlantic_data}
This fact should lower the death rate among infected cases, since our model does not differentiate between different age groups. 

All of these factors point out that the major, if not the only, reason for drastic increases on the number of infections and deaths is the relaxation of the mitigation efforts and by consequence an increase in the infection rate. If the mitigation efforts were to have remained intact, one would naively expect that both the number of infections and fatalities would decrease over time.

We attempt to quantify how many additional people died due to the relaxation of mitigation efforts. We project the number of casualties under the assumption that the mitigation efforts remain intact from the end of April to the end of August, and compare this projection with the actual death counts by the end of August. We choose this date for the reason that the summer peak seems to be passed by the end of August.

\begin{figure}
\centerline{\includegraphics[width=0.45\textwidth]{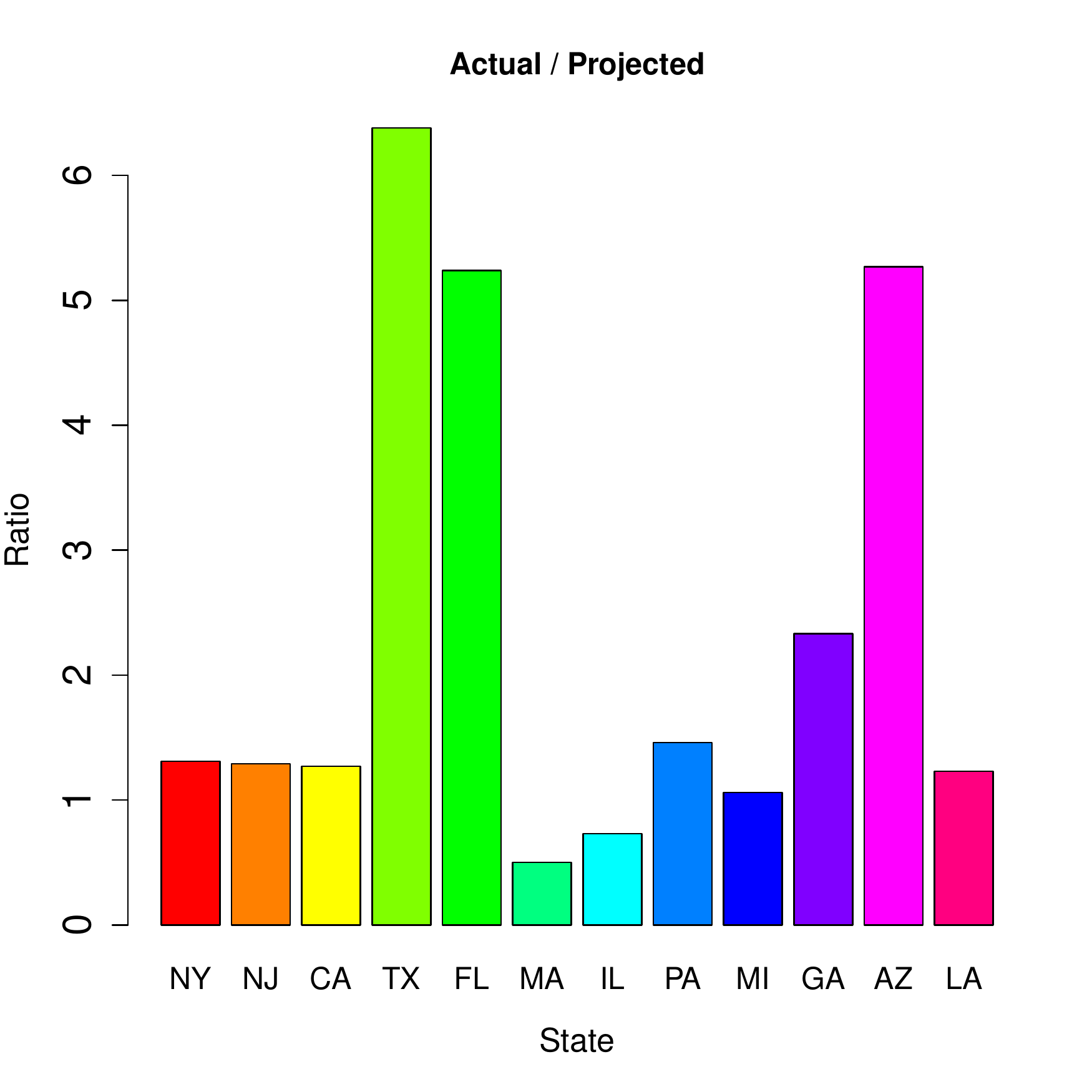}}
\caption{Bar chart with the ratio 
between the actual  and the projected death count by the end of August 
for the twelve states with the largest counts. 
The projected count is calculated assuming that the mitigation efforts 
have been maintained since the end of April. 
This projection represents the ideal upper bound for the  number 
of casualties as other factors which should lower the death count are not taken into account in the present model. See the text for details. 
Most states are within 50\% of the projection. Two states, Massachusetts and Illinois, demonstrate a lower death number relative to the projections. Three states, Texas, Florida, and Arizona, show a substantial increase relative to the projections. Another state which shows an increase greater than a factor of two is Georgia. }
\label{Fig:ratio}
\end{figure}

We investigate the twelve states with the largest number of casualties by the end of August. They are New York, New Jersey, California, Texas, Florida, Massachusetts, Illinois, Pennsylvania, Michigan, Georgia, Arizona, and Louisiana. 
Figs. 1 and 2 display the projected number of deaths, assuming that the mitigation efforts and all of the parameters in the model remain unchanged since the end of April, alongside the actual number of deaths. 
Table I displays the confirmed and the projected number of deaths  by August 31st, as well as the ratio between both counts. Fig. \ref{Fig:ratio} shows the ratio between the actual  and the projected death count by the end of August as a bar chart. 

From the data presented in Figs. 1, 2 and \ref{Fig:ratio}, and Table I, one can conclude that there are marked differences in the increase of the death count among various states. As we discussed above, the major factor in this difference should be due to the distinct policies in relaxing the mitigation efforts. An interesting and important topic is to find any possible correlation between these policies and the increase or decrease of fatalities. 

\begin{center}
\begin{table}
 \begin{tabular}{||l c r c||} 
 \hline
 State & Actual & Projected & Actual/Projected \\ [0.5ex] 
\hline\hline
New York & 32,541 & 24,788 & 1.31 \\
New Jersey & 15,945 & 12,333& 1.29 \\
California & 13,020 & 10,214& 1.27 \\
Texas & 12,857 & 2,015 & 6.38 \\
Florida & 11,186 & 2,133 & 5.24\\
Massachusetts & 9,060&18,184 &0.50 \\
Illinois & 8,258& 11,296 & 0.73 \\
Pennsylvania &7,743 &5,304& 1.46 \\ 
Michigan &6,756 &6,348& 1.06 \\
Georgia & 5,506 & 2,365 & 2.33\\
Arizona  & 5,031 & 955& 5.27 \\
Louisiana & 4,950 & 4,011 &1.23\\
 \hline 
\end{tabular}
\label{Table:ratios}
\caption{The actual confirmed and the projected number of deaths 
by August 31st. The projected count is obtained by assuming the mitigation efforts in place at the end of April were maintained by the end of August.
The last column is the factor of increase (if larger than 1) or decrease (if smaller than 1) of the actual death count when compared with the projection. }

\end{table}
\end{center}

\section{Reopening Schedules of Different States}

In this section, we break down the details of the reopening policies of various states by considering the status of five categories of activities. These include the reopening of outdoor restaurants, indoor restaurants, bars, gyms, and stores. The timings for reopening these businesses vary significantly between states. For example, indoor restaurants were allowed to open on April 27th in Georgia, but not until September 1st in New Jersey. Besides the schedule of reopening of different businesses, the face mask mandate can also be an important policy which affects the spreading of the virus, \cite{CDC_mask}
therefore we also considered the dates face mask mandates were enacted. 
The schedule for reopening and the date for the state mask mandate, 
if applicable, are displayed in Table II.

\begin{center}
\begin{table*}
\begin{tabular}{||p{2.2cm}|p{2cm}p{2cm}p{2cm}p{2cm}p{2cm}p{2cm}||}
 \hline
 State & Outdoor Restaurants
 & Indoor Restaurants
 & Bars & Gyms & Stores & Mask Mandate \\ [0.5ex] 
\hline\hline
New York & Jun 22 & Aug 24 & Jun 22 & May 29 & May 29 & May 28 \\
New Jersey & Jun 15 &  Sep 1 & NA & Jul 8 & Jun 15 & Jun 8 \\
California & May 26 & May 26& Jun 12 & Jun 12 & May 26 & Jun 18 \\
Texas & May 1 & May 1 & May 22 & May 6& May 1& NA \\
Florida & May 11 & May 11 & Jun 5 & May 18 & May 11 & NA\\
Massachusetts & Jun 22 & Jun 22 & NA & Jul 6 & Jun 8 & May 18 \\
Illinois & Jun 26 & Jun 26 & Jun 26 & Jul 20 & May 29 & May 29\\
Pennsylvania & May 29 & May 29 & May 29 & May 29 & May 1 &Jul 1 \\ 
Michigan & May 18 & May 18 & May 22 & Jun 10 & May 26 & Jul 14 \\
Georgia & Apr 27 & Apr 27 & Jun 1 & Apr 24 & Apr 24 & NA\\
Arizona  & May 11 & May 11 & May 15 & May 13 & May 4 & NA \\
Louisiana & May 15 & May 15 & Jun 15 & May 15 & May 15 & Jul 11\\
 \hline 
\end{tabular}
\label{Table:policies}
\caption{The reopening policies of different states. These include the dates of reopening of outdoor restaurants, indoor restaurants, bars, gyms, and stores. The date of the state mask mandate is also listed in the last column. An entry of ``NA'' indicates that a statewide reopening of the particular category or mask mandate has not been implemented by the September 1. \cite{WP_reopening}  }

\end{table*}
\end{center}

\begin{center}
\begin{figure*}
\centerline{\includegraphics[width=1\textwidth]{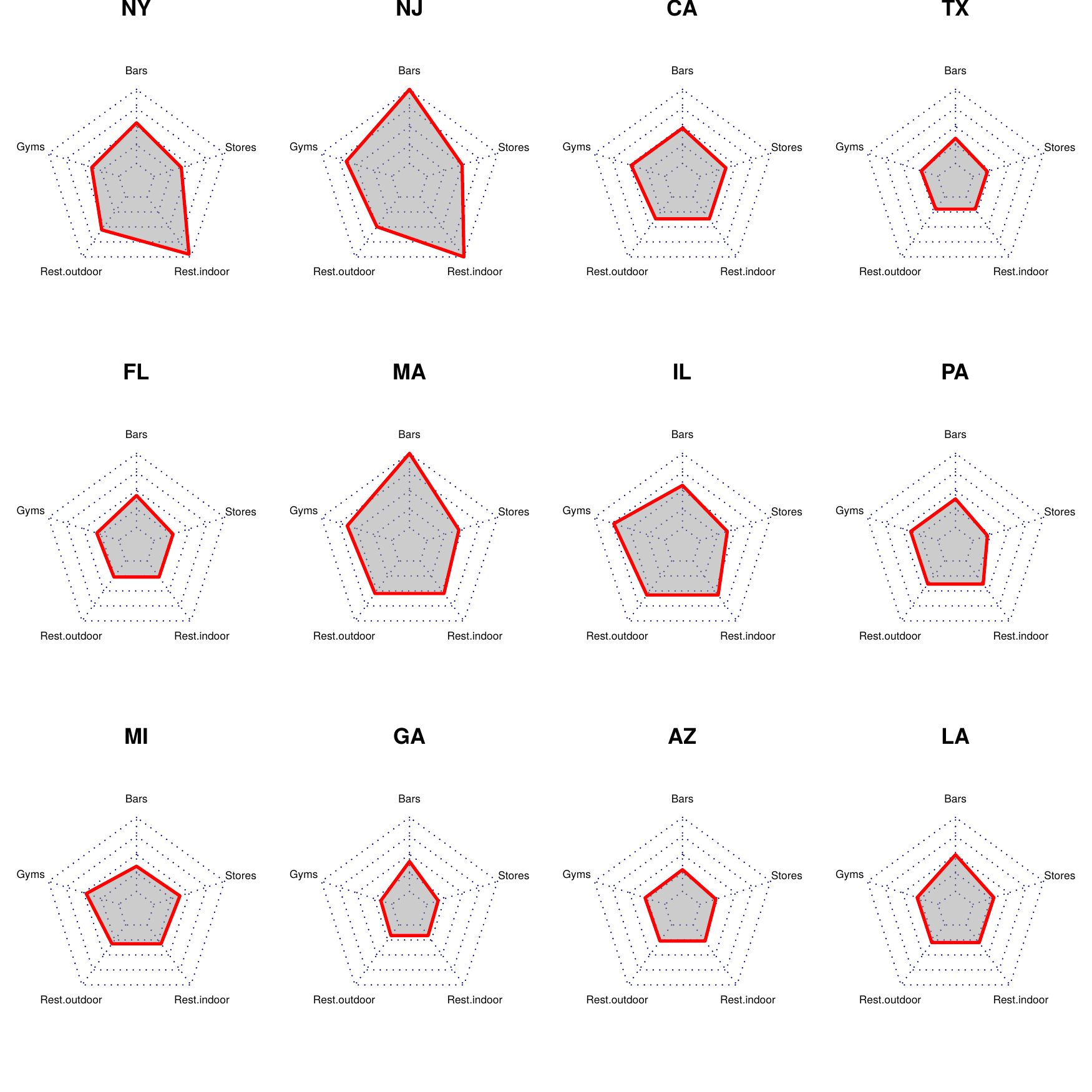}}
\caption{Radar charts for the reopening policies in different states. Five different categories are included: indoor restaurants, outdoor restaurants, gyms, bars, and stores. The earlier the reopening, the closer to the origin of the radar chart the vertex is. The zeroth day (origin of the radar chart) is April 1st, and the last day (the edge of the radar chart) is August 30th. A smaller pentagon implies a more aggressive approach in the reopening.}
\label{fig:radar_chart}
\end{figure*}
\end{center}

For a quicker visual understanding of the reopening policies, we also plot the reopening dates of different categories as a radar chart for all the twelve states we consider
(Fig. \ref{fig:radar_chart}). The size of the pentagon is related to the speed of 
the reopening. The larger the pentagon, the slower the approach to reopening. By contrast, the smaller the pentagon, the more aggressive the reopening policies. 

Previously, we investigated the effects of reopening by considering both the timing of reopening as well as the extent of the prevention measures implemented 
to lower the risk of infection after reopening. We found that lowering the infection rate is more important than the timing of reopening in the long term. \cite{Tam_Walker_Moreno_2020b}
A probable way to lower the infection rate is by wearing personal protective gear, in particular a face mask. \cite{CDC_mask} 

According to the data presented above, the four states (Texas, Florida, Arizona, Georgia) which have the largest increase of deaths due to reopening all have a common feature -- a mask mandate has never been implemented. 
While the reopening date is markedly difference among different states, these four states tend to reopen early, but not by much compared to other states such as Michigan and Louisiana. The reopening dates do not seem to have as strong a correlation as the implementation of a mask mandate to the death count. For comparison, the Pearson correlation coefficients between the reopening date and the increase of the death count from the data presented in the table II fall between 0.5 and 0.7. 

This observation seems to be consistent with the idea that the lowering of infection rate after reopening is more critical than the timing of reopening. \cite{Tam_Walker_Moreno_2020b,Tam_Walker_Moreno_2020a} The reopening dates primarily affect the initial conditions of the epidemic, but the mask mandate affects the dynamics, in particular the infection rate, of the epidemic. By the end of  August, California and Louisiana did not show as substantial an increase in casualties as the above four states, although it is clear that the number of deaths per day is increasing starting around mid-July. 

The two states with a lower death count than projected, Illinois and Massachusetts, both started mask mandates at earlier dates. Together with being comparatively late in reopening, they both show a clear decrease in death counts as one would predict from the trend in April.

While one can legitimately argue that correlation does not imply causation, it does seem that lack of a face mask mandate demonstrates the strongest correlation with the 
increase in the death count after reopening. Delaying the reopening should naively
lower or at least delay the increase in the number of fatalities, but it 
seems to be a secondary factor when comparing to the effect of the mask mandates.
One cannot directly identify the lack of a mask mandate as the cause of
the increase in the number of casualties, as the states which do not have 
a mask mandate may also not have other policies which could help in lowering the infection rate. It is beyond the scope of the present study to uncover those possible causes.

\section{Discussion}

The various mitigation efforts in the spring of 2020, in particular the stay-at-home orders, helped to stabilize the number of infections and casualties.
All states passed or very nearly passed the phase of exponential growth in the number of infections by the end of April. \cite{Tam_Walker_Moreno_2020b} 
Many other factors should have reduced the spread of the virus and the
fatality rate. 
Awareness of the COVID-19 among the public and the general increasing trend of wearing face masks in public areas should contribute to the lowering of the infection count, and thus the death count. Together with substantial improvement in the number of tests administered as compared to last March and the reduction of the share of infections between senior citizens, 
all these factors should reduce the infection and
the death counts.
Thus, if the mitigation efforts had remained intact, the infection rate should have 
decreased over time. Therefore, the hypothetical projection presented in this paper represents an upper bound estimation, since it assumes these factors which potentially reduced the infection rate to be unchanged since the end of April.

The pressure to reopen starting in April led many state governors to relax their states' mitigation efforts. The extent of the relaxation is markedly different between states, as is the change in mortality. \cite{WP_reopening} Our analysis provides a quantitative assessment of the increase in the number of deaths due to the relaxation of the mitigation efforts and as consequence 
does or does not corroborate  the states' policies. This study should provide insight on what is a a good strategy for controlling the infection rate in the process of reopening. While various strategies do incur different impacts on human interactions and consequently the economy and many others, how to balance the effectiveness of lowering the infection rate against the negative effects of reducing human interactions is a broad topic which is beyond the scope of this study. 

From the analysis presented in this paper, we found that three states show the 
largest increase in casualties due to their reopening policies:
Texas, Florida, and Arizona. Georgia also shows substantial increase, though not to as great of an extent. 
On the other hand, Illinois and Massachusetts both show appreciable lowering of the death rate after April.  
A common policy of the four states with the largest increase of deaths 
due to reopening  is the absence of a mask mandate in the state. 
These four states tended to have reopened early, but not excessively so relative to other states.  Therefore, we conclude that  reopening dates do not seem to have correlate as strongly as the implementation of a mask mandate with an increasing number of infections and deaths.

A thorough survey of the differences between policies for states that show a marked increase in the death rate compared to those that show a reduction should provide guidance for a strategy which achieves the best results for lowering the number of infections and deaths. 
In reality, relaxing the mitigation efforts is perhaps not purely a scientific problem, as it involves economic, political, and other social considerations. A delicate balance between all of those factors may need to be found. 

Given that the winter season may present another wave of infections, a thorough study is particularly time sensitive. Our analysis seems to suggest that the face mask mandate is the most important policy for lowering the death count.

\section{Acknowledgment}
We thank Rebecca Christofferson for useful discussions. 
This work use the high performance computational resources provided by the Louisiana Optical Network Initiative (http://www.loni.org) and HPC@LSU computing. 




\end{document}